\documentclass[journal]{IEEEtran}
\usepackage{amsmath,amsthm,amssymb}
\usepackage[colorlinks=true]{hyperref}
\usepackage{algorithm}
\usepackage{algpseudocode}
\usepackage{graphicx,subfigure}
\usepackage{cite}
\usepackage{multirow}

\graphicspath{{pics/}}

\newtheorem{claim}{Claim}

\algblock{ParFor}{EndParFor}
\algnewcommand\algorithmicparfor{\textbf{parfor}}
\algnewcommand\algorithmicpardo{\textbf{do}}
\algnewcommand\algorithmicendparfor{\textbf{end\ parfor}}
\algrenewtext{ParFor}[1]{\algorithmicparfor\ #1\ \algorithmicpardo}
\algrenewtext{EndParFor}{\algorithmicendparfor}

\begin{document}
\title{Undersampled Phase Retrieval via Majorization-Minimization}
\author{Tianyu~Qiu~and~Daniel~P.~Palomar,~\IEEEmembership{Fellow,~IEEE}
\thanks{Tianyu Qiu and Daniel P. Palomar are with The Hong Kong University of Science and Technology (HKUST), Hong Kong (e-mail: tqiu@ust.hk; palomar@ust.hk).}
}



\maketitle

\begin{abstract}
In the undersampled phase retrieval problem, the goal is to recover an $N$-dimensional complex signal $\mathbf{x}$ from only $M<N$ noisy intensity measurements without phase information. This problem has drawn a lot of attention to reduce the number of required measurements since a recent theory established that $M\approx4N$ intensity measurements are necessary and sufficient to recover a generic signal $\mathbf{x}$. In this paper, we propose to exploit the sparsity in the original signal and develop low-complexity algorithms with superior performance based on the majorization-minimization (MM) framework. The proposed algorithms are preferred to existing benchmark methods since at each iteration a simple surrogate problem is solved with a closed-form solution that monotonically decreases the original objective function. Experimental results validate that our algorithms outperform existing up-to-date methods in terms of recovery probability and accuracy, under the same settings.
\end{abstract}


\section{Introduction}
\IEEEPARstart{T}{he} phase retrieval problem is to recover a complex signal $\mathbf{x}\in\mathbb{C}^N$ from the magnitudes of $M$ linear measurements (usually corrupted with noise):
\begin{equation}
\label{eq1.1}
y_i=\left|\mathbf{a}_i^H\mathbf{x}\right|^2+n_i\in\mathbb{R},\;i=1,\ldots,M.
\end{equation}
This problem is motivated by the fact that most of the optical devices (e.g., CCD cameras, human eyes, etc.) can easily measure the intensity, rather than the phase, of the incoming light. Hence, it is a challenging task to uniquely recover the original signal with only intensity information at hand. The measurement vectors $\{\mathbf{a}_i\in\mathbb{C}^N\}_{i=1}^M$ are known beforehand. For example, they correspond to rows of the Discrete Fourier Transform (DFT) matrix in various imaging applications; to name a few, optical imaging \cite{Walther1963}, astronomy \cite{Fienup1987}, crystallography \cite{Harrison1993}, and microscopy \cite{Miao2002}. Other non-imaging applications include audio signal processing \cite{Gerkmann2015} etc. A comprehensive overview on the theory and applications of the phase retrieval problem is discussed in \cite{Shechtman2015}.

The phase retrieval problem is indeed a non-linear non-convex inverse problem. To recover the original signal $\mathbf{x}\in\mathbb{C}^N$ (up to a global phase shift due to the magnitude operator), generally the number of the measurements $M$ should exceed the dimension of the signal $N$. On the theoretical side, the authors in \cite{Candes2013b} have proved that the number of measurements $M$ should at least be $O(N\log N)$ for a successful recovery with high probability when the measurement vectors are chosen independently and uniformly at random on the unit sphere. Furthermore, by designing specific measurement vectors, results have been established that $M\geq4N-O(N)$ measurements are necessary \cite{Heinosaari2013} and $M\geq4N-4$ measurements are sufficient \cite{Bodmann2015} to reconstruct the original signal $\mathbf{x}\in\mathbb{C}^N$ up to a global phase uncertainty. A conjecture has been posed in \cite{Bandeira2014} that $M=4N-4$ generic measurements are both necessary and sufficient. In special cases where $N=2,3$, the conjecture has been proved right. On the algorithmic side, it has been verified empirically that $M\approx4N$ measurements are required to recover the original signal $\mathbf{x}\in\mathbb{C}^N$ with high probability (close to $1$) when the measurement vectors are drawn from independent and identical complex Gaussian distributions \cite{Qiu2016}. As for a real signal, $M\geq2N-1$ measurements have been proved to be both necessary and sufficient to recover $\mathbf{x}\in\mathbb{R}^N$ up to a sign change \cite{Balan2006}. All these results on the minimal number of measurements required to recover the original signal are rather disappointing, especially in applications where only a few measurements are available.

The undersampled phase retrieval problem is to recover an $N$-dimensional complex signal (up to a global phase shift) from only $M<N$ noisy magnitude of its linear measurements. One potential approach is to exploit sparsity in the original signal \cite{Eldar2015,Voroninski2016,Li2013}. And it has been proved that $M\geq8K-2$ measurements are sufficient to recover a $K$-sparse (at most $K$ non-zero elements) complex signal $\mathbf{x}\in\mathbb{C}^N$ under a generic Gaussian measurement vectors setting ($M\geq4K-1$ for real-valued case) \cite{Li2013}. It is quite encouraging that it is possible to recover a signal with fewer intensity measurements than its actual dimension ($M<N$).

Most existing algorithms on undersampled phase retrieval either directly add a constraint to guarantee sparsity \cite{Shechtman2014,Mukherjee2014} or incorporate an additional term in the objective function to promote sparsity \cite{Weller2015,Moravec2007}. The convex $\ell_1$ norm penalty term $\|\mathbf{x}\|_1$ is well-known for producing desired sparse solutions. \cite{Mukherjee2014} and \cite{Moravec2007} share the same idea of incorporating into the classical Fienup algorithm \cite{Fienup1982} a projection step to a set of sparse signals. But \cite{Mukherjee2014} requires the exact number of the sparsity level $K$, and \cite{Moravec2007} requires the exact number of $\|\mathbf{x}\|_1$, which are not available in practice. The authors in \cite{Shechtman2014} combine the damped Gauss-Newton method and the $2$-opt local search method to iteratively update the solution and its support with rough information on the sparsity level $K$. But the combinatorial nature in the support update procedure makes this algorithm impractical when $K$ is relatively large. Besides these, a convex semidefinite programming problem has also been proposed to solve the undersampled phase retrieval problem through the matrix-lifting technique (introducing a new variable $\mathbf{X}:=\mathbf{xx}^H$) \cite{Ohlsson2012}. But the dimension increase in the lifting procedure limits the application to small scale problems. Another different approach is based on the generalized approximate message passing \cite{Schniter2015}. But assumptions on the signal and noise probability distributions may not be accurate. Recently, one robust algorithm (UPRwO) \cite{Weller2015} has been proposed and has been shown to have less reconstruction error than all the former methods for measurements corrupted with outliers and noise.

In this paper, we propose two simple and efficient algorithms based on the majorization-minimization framework to solve the undersampled phase retrieval problem under two different problem settings. The first setting is simple in which the unknown signal is considered to be sparse in the standard basis. The second, more general, setting considers cases where the original signals are not sparse in the standard basis (or any other bases). We share the same idea of applying the sparse coding techniques to the phase retrieval problem \cite{Tillmann2016}, inspired by the fact that a lot of image and video signals can be sparsely approximated by a linear combination of a few columns in a dictionary \cite{Aharon2006,Elad2006,Mairal2010,Mairal2014}. Recently the authors in \cite{Tillmann2016} have shown encouraging results of exploiting sparse coding for the oversampled phase retrieval problem (DOLPHIn algorithm). In this paper, we propose efficient algorithms to recover the unknown signals with high accuracy for the undersampled phase retrieval problem by jointly designing the dictionary and the sparse codes.

The contributions of this paper are:
\begin{enumerate}
	\item Numerical methods for the undersampled phase retrieval problem.
	\item Monotonicity and guarantee of convergence to a stationary point for the sequence of points generated by our algorithms.
	\item Faster numerical convergence of our algorithms compared to state-of-the-art methods UPRwO and DOLPHIn.
	\item Low complexity per iteration of our algorithms (only requiring basic matrix multiplication).
\end{enumerate}

The remaining sections are organized as follows. We first provide a brief introduction on the majorization-minimization framework in section \ref{section2}. Later, we propose an algorithm to solve the undersampled phase retrieval problem of sparse signals using the majorization-minimization techniques. When the unknown signals are not sparse, we propose another algorithm in section \ref{section3} to solve the undersampled phase retrieval problem through sparse coding. Numerical results and comparisons with up-to-date benchmark methods are presented and discussed in section \ref{section4}. Finally, we conclude our work in section \ref{section5}.

\textit{Notation:} Boldface upper case letters (e.g., $\mathbf{X,A}$) denote matrices, while boldface lower case letters (e.g., $\mathbf{x,a}$) denote column vectors, and italics (e.g., $x,a,D$) denote scalars. $\mathbb{R}$ and $\mathbb{C}$ denote the real field and the complex field, respectively. For any complex number $x$, $|x|$ denotes the magnitude, $\arg(x)$ denotes the phase, and $\mathrm{Re}[x]$ denotes the real part. The superscripts $(\cdot)^T$, $\overline{(\cdot)}$, and $(\cdot)^H$ denote transpose, conjugate, and conjugate transpose, respectively. The curled inequality symbol $\succeq$ is used to denote generalized inequality; $\mathbf{a}\succeq\mathbf{b}$ means that $a_i\geq b_i,\forall i$ and $\mathbf{A}\succeq\mathbf{B}$ means that $\mathbf{A}-\mathbf{B}$ is a Hermitian positive semidefinite matrix. $\mathbf{I}_n$ is the $n\times n$ identity matrix (or simply $\mathbf{I}$ when no confusion is caused). $\lambda_{\max}(\mathbf{A})$ denotes the largest eigenvalue of a matrix $\mathbf{A}$. $\odot$ denotes the Hadamard (element-wise) product of two matrices or vectors of the same size. $\mathbf{1}$ is a matrix or vector with all elements $1$, and $\mathbf{0}$ all elements $0$. As usual, $\|\mathbf{x}\|_2$ and $\|\mathbf{x}\|_1$ denote the Euclidean norm and $\ell_1$ norm of a vector $\mathbf{x}$\footnote{In complex-valued case, the absolute operator is replaced by the magnitude operator.}, and $[\mathbf{x}]_i$ denotes its $i$-th element. For a matrix $\mathbf{X}$, $\|\mathbf{X}\|_F$ denotes the Frobenius norm and $\mathbf{X}^\dagger$ denotes its Moore-Penrose pseudoinverse.

\section{Compressive Phase Retrieval via Majorization-Minimization}
\label{section2}
In this section, we first provide a brief overview of the general majorization-minimization (MM) framework. Later, we propose a simple iterative algorithm to solve the undersampled phase retrieval problem of sparse signals via the MM techniques.
\subsection{The MM Algorithm}
The majorization-minimization (MM) algorithm \cite{Hunter2004,Sun2016} is an iterative optimization method, which includs the well-known expectation-maximization (EM) algorithm as a special case. Instead of solving the original difficult optimization problem directly, an MM algorithm deals with a sequence of simple surrogate problems that produce a sequence of points to drive the original objective function downhill.

For a real valued function $f(\boldsymbol{\theta})$, any function $g(\boldsymbol{\theta}\mid\boldsymbol{\theta}^{(m)})$ satisfying the following two conditions is a majorization function of $f(\boldsymbol{\theta})$ at the point $\boldsymbol{\theta}^{(m)}$:
\begin{equation}
\label{eq2.2}
\begin{aligned}
g(\boldsymbol{\theta}\mid\boldsymbol{\theta}^{(m)})&\geq f(\boldsymbol{\theta}),\;\forall \boldsymbol{\theta},\\
g(\boldsymbol{\theta}^{(m)}\mid\boldsymbol{\theta}^{(m)})&=f(\boldsymbol{\theta}^{(m)}).
\end{aligned}
\end{equation}
The function $g(\boldsymbol{\theta}\mid\boldsymbol{\theta}^{(m)})$ is a global upper bound of $f(\boldsymbol{\theta})$ and touches it at the point $\boldsymbol{\theta}^{(m)}$. In general, these majorization functions $\{g(\boldsymbol{\theta}\mid\boldsymbol{\theta}^{(m)})\}_m$ are chosen to be convex and much easier to deal with than the original function, which usually is non-convex or non-differentiable.

Initialized by any feasible point $\boldsymbol{\theta}^{(0)}$, an MM algorithm generates a sequence of points $\{\boldsymbol{\theta}^{(m)}\}_m$ according to the updating rule:
\begin{equation}
\label{eq2.3}
\boldsymbol{\theta}^{(m+1)}\in\underset{\boldsymbol{\theta}}{\arg}\min g(\boldsymbol{\theta}\mid\boldsymbol{\theta}^{(m)}).
\end{equation}
This sequence of points $\{\boldsymbol{\theta}^{(m)}\}_m$ has a favorable property of driving the original objective function $f(\boldsymbol{\theta})$ downhill:
\begin{equation}
\label{eq2.4}
\begin{aligned}
f(\boldsymbol{\theta}^{(m+1)})&\leq g(\boldsymbol{\theta}^{(m+1)}\mid\boldsymbol{\theta}^{(m)})\leq g(\boldsymbol{\theta}^{(m)}\mid\boldsymbol{\theta}^{(m)})\\
&=f(\boldsymbol{\theta}^{(m)}).
\end{aligned}
\end{equation}
The first inequality and the third equality come from the definition of the majorization function \eqref{eq2.2}. The second inequality is valid because $\boldsymbol{\theta}^{(m+1)}$ is a minimizer of $g(\boldsymbol{\theta}\mid\boldsymbol{\theta}^{(m)})$ from \eqref{eq2.3}. Therefore, one can find a stationary point for the original problem by solving the surrogate problems instead.

\subsection{C-PRIME}
Assume that we have in hand $M$ noisy measurements $\{y_i\}_{i=1}^M$ as in \eqref{eq1.1}, and $M<N$. Our goal is to recover the original signal $\mathbf{x}\in\mathbb{C}^N$ up to a global phase shift. Two difficulties make this inverse problem a challenging task. First, the intensity measurements are not linear or convex with regard to $\mathbf{x}$ due to the magnitude operator. Second, the number of unknowns ($N$ complex variables) is larger than the number of measurements ($M$ real numbers). This under-complete non-linear non-convex inverse problem usually has more than one solution despite the trivial global phase factor. Hence, we further assume that the original signal is sparse. For simplicity, in this section, we only consider that the original signal is sparse in the standard basis, and a more general and complicated case where the sparsity lies in a dictionary is discussed in the next section.

Instead of using the intensity measurements $\{y_i\}_{i=1}^M$ directly, we decide to use the modulus measurements $\{\sqrt{y_i}\}_{i=1}^M$ (we assume $y_i\geq 0$ otherwise we just discard this measurement). Justification on the advantage of using modulus information $\{\sqrt{y_i}\}_{i=1}^M$ over intensity information $\{y_i\}_{i=1}^M$ is provided in Appendix A. We propose to solve the following problem to balance the importance of minimizing the sum of squared error and utilizing the prior sparsity information of the original signal:
\begin{equation}
\label{eq2.5}
\underset{\mathbf{x}\in\mathbb{C}^N}{\text{minimize}}\quad\sum\limits_{i=1}^{M}\left(\sqrt{y_i}-\left|\mathbf{a}_i^H\mathbf{x}\right|\right)^2+\rho\|\mathbf{x}\|_1.
\end{equation}
The first term is a data fitting term measuring how well the sought signal $\mathbf{x}$ fits the modulus measurements $\{\sqrt{y_i}\}_{i=1}^M$. This value should be comparable to the noise level for a successful recovery. The second term $\|\mathbf{x}\|_1$ is used to promote sparsity in $\mathbf{x}$. And $\rho$ is a regularization parameter to balance the weights between the sum of squared error and sparsity level to produce a desired solution. 

Furthermore, if we define the measurement matrix $\mathbf{A}:=[\mathbf{a}_1,\ldots,\mathbf{a}_M]^H\in\mathbb{C}^{M\times N}$ and stack the modulus measurements $\{\sqrt{y_i}\}_{i=1}^M$ together as a vector $\sqrt{\mathbf{y}}:=[\sqrt{y_1},\ldots,\sqrt{y_M}]^T$, problem \eqref{eq2.5} can be formulated as in the following more compact form:
\begin{equation}
\label{eq2.6}
\underset{\mathbf{x}\in\mathbb{C}^N}{\text{minimize}}\quad\left\|\sqrt{\mathbf{y}}-|\mathbf{Ax}|\right\|_2^2+\rho\|\mathbf{x}\|_1.
\end{equation}
Here the square root operator $\sqrt{\cdot}$ and the magnitude operator $|\cdot|$ are applied element-wise. This problem is not convex because of the magnitude operator. Using the majorization-minimization technique, we propose an efficient method to solve the convex surrogate problems instead. Note that
\begin{equation}
\label{eq2.7}
\left\|\sqrt{\mathbf{y}}-|\mathbf{Ax}|\right\|_2^2=\mathbf{x}^H\mathbf{A}^H\mathbf{Ax}-2\sqrt{\mathbf{y}}^T|\mathbf{Ax}|+const.,
\end{equation}
where $const.$ is a constant independent of the variable $\mathbf{x}$.

\begin{claim}
\label{claim1}
Let $\mathbf{L}$ be an $N\times N$ Hermitian matrix and $\mathbf{M}$ be another $N\times N$ Hermitian matrix such that $\mathbf{M}\succeq\mathbf{L}$. Then for any point $\mathbf{x}_0\in\mathbb{C}^N$, the quadratic function $\mathbf{x}^H\mathbf{Lx}$ is majorized by $\mathbf{x}^H\mathbf{Mx}+2\mathrm{Re}\left[\mathbf{x}^H(\mathbf{L}-\mathbf{M})\mathbf{x}_0\right]+\mathbf{x}_0^H(\mathbf{M}-\mathbf{L})\mathbf{x}_0$ at $\mathbf{x}_0$.
\end{claim}
\begin{IEEEproof}
The claim is valid simply by rearranging the terms in $(\mathbf{x}-\mathbf{x}_0)^H(\mathbf{M}-\mathbf{L})(\mathbf{x}-\mathbf{x}_0)\geq0$, cf. \cite{Song2014}.
\end{IEEEproof}

According to Claim \ref{claim1}, the first term in \eqref{eq2.7} can be majorized as
\begin{equation}
\mathbf{x}^H\mathbf{A}^H\mathbf{Ax}\leq C\mathbf{x}^H\mathbf{x}+2\mathrm{Re}\left[\mathbf{x}^H(\mathbf{A}^H\mathbf{A}-C\mathbf{I})\mathbf{x}^{(k)}\right]+const.,
\end{equation}
for any constant $C\geq\lambda_{\max}(\mathbf{A}^H\mathbf{A})$. Further, 
\begin{equation}
\begin{aligned}
|\mathbf{Ax}|&=\left|(\mathbf{Ax})\odot e^{-j\arg(\mathbf{Ax}^{(k)})}\right|\\
&\succeq\mathrm{Re}\left[(\mathbf{Ax})\odot e^{-j\arg(\mathbf{Ax}^{(k)})}\right],
\end{aligned}
\end{equation}
the second term in \eqref{eq2.7} can be majorized as
\begin{equation}
\label{eq2.10}
\begin{aligned}
-2\sqrt{\mathbf{y}}^T|\mathbf{Ax}|&\leq-2\sqrt{\mathbf{y}}^T\mathrm{Re}\left[(\mathbf{Ax})\odot e^{-j\arg(\mathbf{Ax}^{(k)})}\right]\\
&=-2\mathrm{Re}\left[(\sqrt{\mathbf{y}}\odot e^{-j\arg(\mathbf{Ax}^{(k)})})^T\mathbf{Ax}\right].
\end{aligned}
\end{equation}
Combining these two majorization functions together, the corresponding surrogate problem for \eqref{eq2.6} is
\begin{equation}
\begin{aligned}
\underset{\mathbf{x}\in\mathbb{C}^N}{\text{minimize}}\quad C\mathbf{x}^H\mathbf{x}+2\mathrm{Re}\left[\mathbf{x}^H(\mathbf{A}^H\mathbf{A}-C\mathbf{I})\mathbf{x}^{(k)}\right]\\
-2\mathrm{Re}\left[(\sqrt{\mathbf{y}}\odot e^{-j\arg(\mathbf{Ax}^{(k)})})^T\mathbf{Ax}\right]+\rho\|\mathbf{x}\|_1.
\end{aligned}
\end{equation}
This surrogate problem is convex in $\mathbf{x}$ and is equivalent to the following problem:
\begin{equation}
\label{eq2.12}
\underset{\mathbf{x}\in\mathbb{C}^N}{\text{minimize}}\quad\|\mathbf{x}-\mathbf{c}\|_2^2+\frac{\rho}{C}\|\mathbf{x}\|_1,
\end{equation}
which has a simple closed-form solution
\begin{equation}
\mathbf{x}^{\star}=e^{j\arg(\mathbf{c})}\odot\max\left\{\mathbf{c}-\frac{\rho}{2C}\mathbf{1},\mathbf{0}\right\}.
\end{equation}
The vector $\mathbf{c}$ is a constant independent of the variable $\mathbf{x}$:
\begin{equation}
\label{eq2.14}
\mathbf{c}:=\mathbf{x}^{(k)}-\frac{1}{C}\mathbf{A}^H\left(\mathbf{A}\mathbf{x}^{(k)}-\sqrt{\mathbf{y}}\odot e^{j\arg(\mathbf{Ax}^{(k)})}\right).
\end{equation}

Now it is clear to see the benefits of using the majorization-minimization framework. Instead of dealing with the original non-convex non-differentiable problem \eqref{eq2.6}, we only need to solve a surrogate problem \eqref{eq2.12} which has a simple closed-form solution at every iteration. We name our algorithm compressive phase retrieval via the majorization-minimization technique (C-PRIME for short) and summarize the procedure in Algorithm \ref{alg1}.

In the algorithm, we further adopt the SQUAREM algorithm \cite{VARADHAN2008} to accelerate the convergence speed of our method. SQUAREM generally achieves a superlinear convergence rate and only requires parameter updating. Instead of updating $\mathbf{x}^{(k+1)}$ directly from $\mathbf{x}^{(k)}$ at the $k$-th iteration, SQUAREM first seeks an intermediate point $\mathbf{x}_3$ based on $\mathbf{x}^{(k)}$ and later updates the next point $\mathbf{x}^{(k+1)}$ from this intermediate point. Unfortunately, this updating rule may violate the descent property of the MM framework. Therefore, we add a backtracking step in our algorithm (the while loop) to maintain the descent property. In detail, we repeatedly halve the distance between $\alpha$ and $-1$ until the descent property is valid. This strategy is guaranteed to work because in the worst case where $\alpha=-1$, the intermediate point satisfies $\mathbf{x}_3=\mathbf{x}^{(k)}+2\mathbf{r}+\mathbf{v}=\mathbf{x}_2$, which ensures that the algorithm will jump out of the while loop. (Actually it only takes several updates $\alpha\leftarrow(\alpha-1)/2$ for the descent property to be maintained in the simulation.)

\begin{algorithm}
	\renewcommand{\algorithmicrequire}{\textbf{Input:}}
	\renewcommand{\algorithmicensure}{\textbf{Output:}}
	\caption{C-PRIME}
	\label{alg1}
	\begin{algorithmic}[1]
		\Require $\mathbf{A},\mathbf{y},\rho,t_0$ (maximum iteration number)
		\State Initial $\mathbf{x}^{(0)}\leftarrow$ Random vector
		\State Choose a constant $C\geq\lambda_{\max}(\mathbf{A}^H\mathbf{A})$
		\For {$k=0,\ldots,t_0-1$}
			\State $\mathbf{c}_1=\mathbf{x}^{(k)}-\frac{1}{C}\mathbf{A}^H\left(\mathbf{A}\mathbf{x}^{(k)}-\sqrt{\mathbf{y}}\odot e^{j\arg(\mathbf{Ax}^{(k)})}\right)$
			\State $\mathbf{x}_1=e^{j\arg(\mathbf{c}_1)}\odot\max\left\{\mathbf{c}_1-\frac{\rho}{2C}\mathbf{1},\mathbf{0}\right\}$
			\State $\mathbf{c}_2=\mathbf{x}_1-\frac{1}{C}\mathbf{A}^H\left(\mathbf{A}\mathbf{x}_1-\sqrt{\mathbf{y}}\odot e^{j\arg(\mathbf{Ax}_1)}\right)$
			\State $\mathbf{x}_2=e^{j\arg(\mathbf{c}_2)}\odot\max\left\{\mathbf{c}_2-\frac{\rho}{2C}\mathbf{1},\mathbf{0}\right\}$
			\State $\mathbf{r}=\mathbf{x}_1-\mathbf{x}^{(k)}$
			\State $\mathbf{v}=\mathbf{x}_2-\mathbf{x}_1-\mathbf{r}$
			\State $\alpha\leftarrow-\frac{\|\mathbf{r}\|_2}{\|\mathbf{v}\|_2}$
			\State $\mathbf{x}_3\leftarrow\mathbf{x}^{(k)}-2\alpha\mathbf{r}+\alpha^2\mathbf{v}$
			\While{$\|\sqrt{\mathbf{y}}-\left|\mathbf{Ax}_3\right|\|_2^2+\rho\|\mathbf{x}_3\|_1>\|\sqrt{\mathbf{y}}-\left|\mathbf{Ax}_2\right|\|_2^2+\rho\|\mathbf{x}_2\|_1$}
				\State $\alpha\leftarrow(\alpha-1)/2$
				\State $\mathbf{x}_3\leftarrow\mathbf{x}^{(k)}-2\alpha\mathbf{r}+\alpha^2\mathbf{v}$
			\EndWhile
			\State $\mathbf{c}_3=\mathbf{x}_3-\frac{1}{C}\mathbf{A}^H\left(\mathbf{A}\mathbf{x}_3-\sqrt{\mathbf{y}}\odot e^{j\arg(\mathbf{Ax}_3)}\right)$
			\State $\mathbf{x}^{(k+1)}=e^{j\arg(\mathbf{c}_3)}\odot\max\left\{\mathbf{c}_3-\frac{\rho}{2C}\mathbf{1},\mathbf{0}\right\}$
		\EndFor
		\Ensure $\mathbf{x}^{(t_0)}$.
	\end{algorithmic}
\end{algorithm}

\section{Sparse Coding for Phase Retrieval}
\label{section3}
In the last section, we proposed a simple and efficient MM-based algorithm to solve the undersampled phase retrieval problem for signals that are sparse in the standard basis. But what if the original signal is only sparse with regard to another (known or unknown) basis\footnote{Vector $\mathbf{x}=\mathbf{\Phi}\mathbf{z}$ where $\mathbf{\Phi}\in\mathbb{C}^{N\times N}$ is a basis and $\mathbf{z}\in\mathbb{C}^N$ is a sparse vector.}, or more generally a dictionary\footnote{Vector $\mathbf{x}=\mathbf{Dz}$ where $\mathbf{D}\in\mathbb{C}^{N\times L}$ is a dictionary and $\mathbf{z}\in\mathbb{C}^L$ is a sparse vector. Matrix $\mathbf{D}$ is named a dictionary in the sense that $\mathbf{x}$ can be represented as a linear combination of the columns in $\mathbf{D}$.}? Typical examples include many image processing applications (image compression, video coding, etc.) where the target images are not sparse in the image domain, but instead are sparse in a transform (e.g., discrete cosine transform or wavelet transform) domain. Besides these, the authors in \cite{Aharon2006,Elad2006} have shown the advantage of learning an overcomplete dictionary to sparsely represent a signal in an image denoising application. Recently, the dictionary learning techniques \cite{Mairal2010,Mairal2014} have also been exploited to solve the phase retrieval problem \cite{Tillmann2016}. Inspired by these sparse coding ideas, we propose a simple and efficient algorithm in this section to solve the undersampled phase retrieval problem for signals that are not sparse in the standard basis.

The same as in the sparse signal case discussed in the last section, it is impossible to uniquely recover the original signal with only undersampled intensity measurements and no additional information provided. In this section, instead of the rigid assumption of sparsity in the standard basis, we only assume that the original signal $\mathbf{x}\in\mathbb{C}^N$ admits a sparse approximation over an unknown overcomplete dictionary $\mathbf{D}\in\mathbb{C}^{N\times L}$ with $L>N$, i.e., $\mathbf{x}\approx\mathbf{Dz}$ and $\mathbf{z}\in\mathbb{C}^L$ is the sparse coding vector. Each column $\mathbf{d}_l$ in the dictionary $\mathbf{D}$ is called an atom and is restricted to be in the unit ball $\|\mathbf{d}_l\|_2\leq1$. Now since the unknown original signal is assumed to be approximated by a linear combination of a few atoms in an unknown dictionary, the following problem is considered to jointly recover the signal and design the dictionary:
\begin{equation}
\label{eq3.15}
\begin{aligned}
&\underset{\mathbf{x},\mathbf{D},\mathbf{z}}{\text{minimize}}&&\|\sqrt{\mathbf{y}}-|\mathbf{Ax}|\|_2^2+\mu\|\mathbf{x}-\mathbf{Dz}\|_2^2+\rho\|\mathbf{z}\|_1\\
&\text{subject to}&&\mathbf{D}\in\mathcal{D},
\end{aligned}
\end{equation}
where $\mathcal{D}$ is a convex set defined as
\begin{equation}
\mathcal{D}:=\{\mathbf{D}\in\mathbb{C}^{N\times L}\mid\|\mathbf{d}_l\|_2\leq 1,\forall l=1,\ldots,L\}.
\end{equation}
The data fitting term in the objective is the same as in the sparse signal case discussed in the last section, the second term measures how well the unknown signal can be approximated by the dictionary, and the last term promotes sparse code so that only a few atoms are chosen to approximate the unknown signal. The two regularization parameters $\mu$ and $\rho$ are used to balance the weights on the data fitting, the dictionary representation, and the sparse code. Unfortunately, there is more than one solution for problem \eqref{eq3.15} because the unknown dictionary is considered as an additional variable. Now that one single signal is insufficient to uniquely determine the dictionary, multiple signals should be exploited to jointly recover the original signals and unknown dictionary.

Assume there are multiple (independent) unknown signals $\{\mathbf{x}_p\in\mathbb{C}^N\}_{p=1}^P$, (we can divide a high
dimensional signal into several equal-length signals of smaller dimensions, e.g., divide a large image into several small patches), and for every signal $\mathbf{x}_p$, we only have a few undersampled (noisy) intensity measurements $\mathbf{y}_p\in\mathbb{R}^M$ ($M<N$). Furthermore, each (independent) signal $\mathbf{x}_p$ is assumed to be sparsely approximated by a linear combination of a few atoms in a shared unknown overcomplete dictionary $\mathbf{D}\in\mathbb{C}^{N\times L}$; $\mathbf{x}_p\approx\mathbf{Dz}_p$. We propose to solve the following problem to recover the multiple original signals from only undersampled intensity measurements:
\begin{equation}
\label{eq3.17}
\begin{aligned}
&\underset{\{\mathbf{x}_p\},\mathbf{D},\{\mathbf{z}_p\}}{\text{minimize}}&&\sum\limits_{p=1}^{P}\left(\|\sqrt{\mathbf{y}_p}-|\mathbf{Ax}_p|\|_2^2+\mu\|\mathbf{x}_p-\mathbf{Dz}_p\|_2^2\right.\\
&&&\left.+\rho\|\mathbf{z}_p\|_1\right)\\
&\text{subject to}&&\mathbf{D}\in\mathcal{D}.
\end{aligned}
\end{equation}
The number of atoms should be less than the number of unknown signals $L<P$. Otherwise, each signal is trivially represented by a $1$-sparse vector $\mathbf{z}_p$ after including $\mathbf{x}_p/\|\mathbf{x}_p\|_2$ as an atom in the dictionary.

Problem \eqref{eq3.17} is not convex, not only because of the magnitude operator, but also because of the quadratic term $\mathbf{Dz}_p$. But the problem is convex with regard to $\mathbf{D}$ if $\{\mathbf{x}_p\}$ and $\{\mathbf{z}_p\}$ are fixed. Also, it is convex with regard to $\{\mathbf{z}_p\}$ when $\{\mathbf{x}_p\}$ and $\mathbf{D}$ are fixed. Another problem is that all variables are tangled together because of the shared dictionary $\mathbf{D}$. But once $\mathbf{D}$ is fixed, problem \eqref{eq3.17} can be separated into $P$ independent smaller problems. Therefore, we propose to solve this problem using the block successive upper-bound minimization method (BSUM) \cite{Hong2016}. BSUM is a simple and iterative algorithm framework to successively optimize upper bounds functions, instead of the original objective function, in a block by block manner. And the convergence analysis is provided in \cite{Razaviyayn2013}.

\subsection{Updating the Sparse Codes $\{\mathbf{z}_p\}$}
We first consider updating $\{\mathbf{z}_p^{(k+1)}\}$ at the $k+1$-th iteration. When $\{\mathbf{x}_p\}$ and $\mathbf{D}$ are fixed to be $\{\mathbf{x}_p^{(k)}\}$ and $\mathbf{D}^{(k)}$, problem \eqref{eq3.17} is equivalent to the following problem:
\begin{equation}
\underset{\{\mathbf{z}_p\}}{\text{minimize}}\quad\sum\limits_{p=1}^{P}\left(\|\mathbf{D}^{(k)}\mathbf{z}_p-\mathbf{x}_p^{(k)}\|_2^2+\frac{\rho}{\mu}\|\mathbf{z}_p\|_1\right),
\end{equation}
which can be separated into $P$ independent problems in the form:
\begin{equation}
\label{eq3.19}
\underset{\mathbf{z}_p}{\text{minimize}}\quad\|\mathbf{D}^{(k)}\mathbf{z}_p-\mathbf{x}_p^{(k)}\|_2^2+\frac{\rho}{\mu}\|\mathbf{z}_p\|_1.
\end{equation}
This is a typical sparse coding problem \cite{Mairal2014}. If the dictionary $\mathbf{D}^{(k)}$ satisfies $(\mathbf{D}^{(k)})^H\mathbf{D}^{(k)}=\mathbf{I}$ (semi-unitary), \eqref{eq3.19} is equivalent to the following problem:
\begin{equation}
\underset{\mathbf{z}_p}{\text{minimize}}\quad\|\mathbf{z}_p-(\mathbf{D}^{(k)})^H\mathbf{x}_p^{(k)}\|_2^2+\frac{\rho}{\mu}\|\mathbf{z}_p\|_1,
\end{equation}
which has a simple closed-form solution
\begin{equation}
\mathbf{z}_p^{\star}=e^{j\arg\left((\mathbf{D}^{(k)})^H\mathbf{x}_p^{(k)}\right)}\odot\max\left\{(\mathbf{D}^{(k)})^H\mathbf{x}_p^{(k)}-\frac{\rho}{2\mu}\mathbf{1},\mathbf{0}\right\}.
\end{equation}
When $(\mathbf{D}^{(k)})^H\mathbf{D}^{(k)}\neq\mathbf{I}$, it is difficult to find a simple closed-form solution for \eqref{eq3.19} directly. Hence we propose to solve a surrogate problem instead. Using the same technique in Claim \ref{claim1}, the first term in \eqref{eq3.19} can be majorized as
\begin{equation}
\begin{aligned}
&\|\mathbf{D}^{(k)}\mathbf{z}_p-\mathbf{x}_p^{(k)}\|_2^2\\
=&\mathbf{z}_p^H(\mathbf{D}^{(k)})^H\mathbf{D}^{(k)}\mathbf{z}_p-2\mathrm{Re}\left[(\mathbf{x}_p^{(k)})^H\mathbf{D}^{(k)}\mathbf{z}_p\right]+const.\\
\leq&E^{(k)}\mathbf{z}_p^H\mathbf{z}_p+2\mathrm{Re}\left[\mathbf{z}_p^H\left((\mathbf{D}^{(k)})^H\mathbf{D}^{(k)}-E^{(k)}\mathbf{I}\right)\mathbf{z}_p^{(k)}\right]\\
&-2\mathrm{Re}\left[(\mathbf{x}_p^{(k)})^H\mathbf{D}^{(k)}\mathbf{z}_p\right]+const.\\
=&E^{(k)}\|\mathbf{z}_p-\mathbf{e}_p\|_2^2+const.,
\end{aligned}
\end{equation}
where $\mathbf{e}_p$ is independent of the variable $\mathbf{z}_p$ and is defined as 
\begin{equation}
\mathbf{e}_p:=\mathbf{z}_p^{(k)}-\frac{1}{E^{(k)}}(\mathbf{D}^{(k)})^H\left(\mathbf{D}^{(k)}\mathbf{z}_p^{(k)}-\mathbf{x}_p^{(k)}\right).
\end{equation}
The scalar $E^{(k)}$ is a constant larger than $\lambda_{\max}\left((\mathbf{D}^{(k)})^H\mathbf{D}^{(k)}\right)$ and we prove in Appendix B that $E^{(k)}\geq L$ is sufficient for a valid majorization function. Therefore, the corresponding surrogate problem for \eqref{eq3.19} is
\begin{equation}
\underset{\mathbf{z}_p}{\text{minimize}}\quad\|\mathbf{z}_p-\mathbf{e}_p\|_2^2+\frac{\rho}{E^{(k)}\mu}\|\mathbf{z}_p\|_1,
\end{equation}
and it has a simple closed-form solution
\begin{equation}
\mathbf{z}_p^{\star}=e^{j\arg(\mathbf{e}_p)}\odot\max\left\{\mathbf{e}_p-\frac{\rho}{2E^{(k)}\mu}\mathbf{1},\mathbf{0}\right\}.
\end{equation}

\subsection{Updating the Estimated Signals $\{\mathbf{x}_p\}$}
When $\mathbf{D}$ and $\{\mathbf{z}_p\}$ are fixed to be $\mathbf{D}^{(k)}$ and $\{\mathbf{z}_p^{(k+1)}\}$, updating $\{\mathbf{x}_p\}$ leads to solving the following problem:
\begin{equation}
\underset{\{\mathbf{x}_p\}}{\text{minimize}}\quad\sum\limits_{p=1}^{P}\left(\|\sqrt{\mathbf{y}_p}-|\mathbf{Ax}_p|\|_2^2+\mu\|\mathbf{x}_p-\mathbf{D}^{(k)}\mathbf{z}_p^{(k+1)}\|_2^2\right),
\end{equation}
which also can be separated into $P$ independent problems:
\begin{equation}
\label{eq3.27}
\underset{\mathbf{x}_p}{\text{minimize}}\quad\|\sqrt{\mathbf{y}_p}-|\mathbf{Ax}_p|\|_2^2+\mu\|\mathbf{x}_p-\mathbf{D}^{(k)}\mathbf{z}_p^{(k+1)}\|_2^2.
\end{equation}
This problem is not convex due to the magnitude operator. Hence, we choose to solve a surrogate problem instead. According to Claim \ref{claim1} and \eqref{eq2.10}, choosing a constant $F\geq\lambda_{\max}(\mathbf{A}^H\mathbf{A})$, the objective function can be majorized as
\begin{equation}
\begin{aligned}
&\|\sqrt{\mathbf{y}_p}-|\mathbf{Ax}_p|\|_2^2+\mu\|\mathbf{x}_p-\mathbf{D}^{(k)}\mathbf{z}_p^{(k+1)}\|_2^2\\
=&\mathbf{x}_p^H\mathbf{A}^H\mathbf{A}\mathbf{x}_p-2\sqrt{\mathbf{y}_p}^T|\mathbf{Ax}_p|\\
&+\mu\mathbf{x}_p^H\mathbf{x}_p-2\mu\mathrm{Re}\left[\mathbf{x}_p^H\mathbf{D}^{(k)}\mathbf{z}_p^{(k+1)}\right]+const.\\
\leq&F\mathbf{x}_p^H\mathbf{x}_p+2\mathrm{Re}\left[\mathbf{x}_p^H(\mathbf{A}^H\mathbf{A}-F\mathbf{I})\mathbf{x}_p^{(k)}\right]\\
&-2\mathrm{Re}\left[(\sqrt{\mathbf{y}_p}\odot e^{-j\arg(\mathbf{Ax}_p^{(k)})})^T\mathbf{Ax}_p\right]\\
&+\mu\mathbf{x}_p^H\mathbf{x}_p-2\mu\mathrm{Re}\left[\mathbf{x}_p^H\mathbf{D}^{(k)}\mathbf{z}_p^{(k+1)}\right]+const.\\
=&(F+\mu)\|\mathbf{x}_p-\mathbf{f}_p\|_2^2+const.,
\end{aligned}
\end{equation}
where $\mathbf{f}_p$ is a constant with regard to the variable $\mathbf{x}_p$:
\begin{equation}
\begin{aligned}
\mathbf{f}_p=&\frac{1}{F+\mu}\left[F\mathbf{x}_p^{(k)}-\mathbf{A}^H\left(\mathbf{A}\mathbf{x}_p^{(k)}-\sqrt{\mathbf{y}_p}\odot e^{j\arg(\mathbf{Ax}_p^{(k)})}\right)\right.\\
&\left.+\mu\mathbf{D}^{(k)}\mathbf{z}_p^{(k+1)}\right].
\end{aligned}
\end{equation}
Therefore, the surrogate problem for \eqref{eq3.27} is
\begin{equation}
\underset{\mathbf{x}_p}{\text{minimize}}\quad\|\mathbf{x}_p-\mathbf{f}_p\|_2^2,
\end{equation}
and it has a simple closed-form solution
\begin{equation}
\mathbf{x}_p^{\star}=\mathbf{f}_p.
\end{equation}

Note that this constant $\mathbf{f}_p$ is similar to the constant $\mathbf{c}$ in \eqref{eq2.14}, which was used to update the signal in the last section. The additional third term $\mathbf{D}^{(k)}\mathbf{z}_p^{(k+1)}$ in $\mathbf{f}_p$ is due to the second approximation over a dictionary term in \eqref{eq3.17}.

\subsection{Updating the Dictionary $\mathbf{D}$}
The final step is to update the dictionary $\mathbf{D}$ fixing the other two blocks of variables $\{\mathbf{x}_p\}$ and $\{\mathbf{z}_p\}$ as $\{\mathbf{x}_p^{(k+1)}\}$ and $\{\mathbf{z}_p^{(k+1)}\}$. Since the regularization parameter $\mu\geq0$, we need to solve the following problem:
\begin{equation}
\begin{aligned}
&\underset{\mathbf{D}}{\text{minimize}}&&\sum\limits_{p=1}^{P}\|\mathbf{x}_p^{(k+1)}-\mathbf{Dz}_p^{(k+1)}\|_2^2\\
&\text{subject to}&&\mathbf{D}\in\mathcal{D},
\end{aligned}
\end{equation}
which can be formulated in a more compact form:
\begin{equation}
\begin{aligned}
&\underset{\mathbf{D}}{\text{minimize}}&&\|\mathbf{X}^{(k+1)}-\mathbf{DZ}^{(k+1)}\|_F^2\\
&\text{subject to}&&\mathbf{D}\in\mathcal{D},
\end{aligned}
\end{equation}
where $\mathbf{X}^{(k+1)}:=[\mathbf{x}_1^{(k+1)},\ldots,\mathbf{x}_P^{(k+1)}]\in\mathbb{C}^{N\times P}$ and $\mathbf{Z}^{(k+1)}:=[\mathbf{z}_1^{(k+1)},\ldots,\mathbf{z}_P^{(k+1)}]\in\mathbb{C}^{L\times P}$. This problem is convex but it is not easy to find a closed-form solution. Matrix $\mathbf{X}^{(k+1)}(\mathbf{Z}^{(k+1)})^H(\mathbf{Z}^{(k+1)}(\mathbf{Z}^{(k+1)})^H)^{-1}$ is a solution if it does not have the constraint $\mathbf{D}\in\mathcal{D}$ and assuming $\mathbf{Z}^{(k+1)}(\mathbf{Z}^{(k+1)})^H$ is invertible. In this subsection, we find a simple closed-form solution for every column $\mathbf{d}_l$ in the dictionary. Instead of treating $\mathbf{D}$ as one variable, we can divide it into $L$ different variables $\{\mathbf{d}_l\}_{l=1}^L$ and update them one by one while keeping the others fixed. To update the $l$-th column $\mathbf{d}_l$, we only need to solve the following problem:
\begin{equation}
\label{eq3.34}
\begin{aligned}
&\underset{\mathbf{d}_l}{\text{minimize}}&&\|\mathbf{X}^{(k+1)}-\sum\limits_{m\neq l}\mathbf{d}_m^{(k)}\mathbf{z}_{m,T}^{(k+1)}-\mathbf{d}_l\mathbf{z}_{l,T}^{(k+1)}\|_F^2\\
&\text{subject to}&&\|\mathbf{d}_l\|_2\leq1,
\end{aligned}
\end{equation}
where $\mathbf{z}_{l,T}^{(k+1)}$ is a row vector denoting the $l$-th row in matrix $\mathbf{Z}^{(k+1)}$. The objective function in \eqref{eq3.34} satisfies
\begin{equation}
\begin{aligned}
&\|\mathbf{X}^{(k+1)}-\sum\limits_{m\neq l}\mathbf{d}_m^{(k)}\mathbf{z}_{m,T}^{(k+1)}-\mathbf{d}_l\mathbf{z}_{l,T}^{(k+1)}\|_F^2\\
=&\|\mathbf{X}^{(k+1)}-\mathbf{D}^{(k)}\mathbf{Z}^{(k+1)}+\mathbf{d}_l^{(k)}\mathbf{z}_{l,T}^{(k+1)}-\mathbf{d}_l\mathbf{z}_{l,T}^{(k+1)}\|_F^2\\
=&\|\mathbf{d}_l\|_2^2\|\mathbf{z}_{l,T}^{(k+1)}\|_2^2-2\mathrm{Re}\left[\mathbf{d}_l^H\left(\mathbf{X}^{(k+1)}-\mathbf{D}^{(k)}\mathbf{Z}^{(k+1)}\right.\right.\\
&\left.\left.+\mathbf{d}_l^{(k)}\mathbf{z}_{l,T}^{(k+1)}\right)(\mathbf{z}_{l,T}^{(k+1)})^H\right]+const.\\
=&\|\mathbf{z}_{l,T}^{(k+1)}\|_2^2\|\mathbf{d}_l-\mathbf{g}_l\|_2^2+const.,
\end{aligned}
\end{equation}
where $\mathbf{g}_l$ is a constant with regard to the variable $\mathbf{d}_l$:
\begin{equation}
\mathbf{g}_l:=\mathbf{d}_l^{(k)}+\frac{1}{\|\mathbf{z}_{l,T}^{(k+1)}\|_2^2}\left(\mathbf{X}^{(k+1)}-\mathbf{D}^{(k)}\mathbf{Z}^{(k+1)}\right)(\mathbf{z}_{l,T}^{(k+1)})^H.
\end{equation}
So problem \eqref{eq3.34} is equivalent to
\begin{equation}
\begin{aligned}
&\underset{\mathbf{d}_l}{\text{minimize}}&&\|\mathbf{d}_l-\mathbf{g}_l\|_2^2\\
&\text{subject to}&&\|\mathbf{d}_l\|_2\leq1,
\end{aligned}
\end{equation}
which has a simple closed-form solution
\begin{equation}
\mathbf{d}_l^{\star}=\frac{\mathbf{g}_l}{\max\{\|\mathbf{g}_l\|_2,1\}}.
\end{equation}
Note that we only need to calculate $\mathbf{X}^{(k+1)}-\mathbf{D}^{(k)}\mathbf{Z}^{(k+1)}$ once to update $\{\mathbf{d}_l\}_{l=1}^L$ in parallel.

Finally, the procedures are summarized in Algorithm \ref{alg2}. We name the algorithm sparse coding for phase retrieval via the majorization-minimization technique (SC-PRIME for short). To make the algorithm concise, the steps of updating the sparse codes $\{\mathbf{z}_p\}$ and the estimated signals $\{\mathbf{x}_p\}$ are written in matrix form. To start, the estimated signal $\mathbf{X}^{(0)}$ and the dictionary $\mathbf{D}^{(0)}$ are initialized as random matrices (the columns of $\mathbf{D}^{(0)}$ are scaled if necessary to ensure the feasibility $\mathbf{D}^{(0)}\in\mathcal{D}$). The initial sparse codes $\mathbf{Z}^{(0)}$ are set as the value minimizing $\|\mathbf{X}^{(0)}-\mathbf{D}^{(0)}\mathbf{Z}\|_F^2$, the matrix $\mathbf{Y}$ in step $9$ is defined as $\mathbf{Y}:=[\mathbf{y}_1,\ldots,\mathbf{y}_P]\in\mathbb{R}^{M\times P}$, and the columns in the dictionary are updated in parallel.

\begin{algorithm}
	\renewcommand{\algorithmicrequire}{\textbf{Input:}}
	\renewcommand{\algorithmicensure}{\textbf{Output:}}
	\caption{SC-PRIME}
	\label{alg2}
	\begin{algorithmic}[1]
		\Require $\mathbf{A},\mathbf{Y},\mu,\rho,t_0$ (maximum iteration number)
		\State Initial $\mathbf{X}^{(0)}\leftarrow$ Random matrix
		\State Initial $\mathbf{D}^{(0)}\leftarrow$ Random matrix ($\|\mathbf{d}_l^{(0)}\|_2\leq1$)
		\State Innitial $\mathbf{Z}^{(0)}\leftarrow\left((\mathbf{D}^{(0)})^H\mathbf{D}^{(0)}\right)^{\dagger}(\mathbf{D}^{(0)})^H\mathbf{X}^{(0)}$
		\State Choose a constant $F\geq\lambda_{\max}(\mathbf{A}^H\mathbf{A})$
		\For {$k=0,\ldots,t_0-1$}
		\State Choose a constant $E^{(k)}\geq\lambda_{\max}\left((\mathbf{D}^{(k)})^H\mathbf{D}^{(k)}\right)$
		\State $\mathbf{E}\leftarrow\mathbf{Z}^{(k)}-\frac{1}{E^{(k)}}(\mathbf{D}^{(k)})^H\left(\mathbf{D}^{(k)}\mathbf{Z}^{(k)}-\mathbf{X}^{(k)}\right)$
		\State $\mathbf{Z}^{(k+1)}=e^{j\arg(\mathbf{E})}\odot\max\left\{\mathbf{E}-\frac{\rho}{2E^{(k)}\mu}\mathbf{1},\mathbf{0}\right\}.$
		\State $\begin{aligned}\mathbf{X}^{(k+1)}&=\frac{1}{F+\mu}\left[F\mathbf{X}^{(k)}-\mathbf{A}^H\left(\mathbf{A}\mathbf{X}^{(k)}\right.\right.\\
		&\left.\left.-\sqrt{\mathbf{Y}}\odot e^{j\arg(\mathbf{AX}^{(k)})}\right)+\mu\mathbf{D}^{(k)}\mathbf{Z}^{(k+1)}\right]\end{aligned}$	
		\ParFor {$l=1,\ldots,L$}
		\State $\mathbf{g}_l\leftarrow\mathbf{d}_l^{(k)}+\frac{\left(\mathbf{X}^{(k+1)}-\mathbf{D}^{(k)}\mathbf{Z}^{(k+1)}\right)(\mathbf{z}_{l,T}^{(k+1)})^H}{\|\mathbf{z}_{l,T}^{(k+1)}\|_2^2}$
		\State $\mathbf{d}_l^{(k+1)}=\frac{\mathbf{g}_l}{\max\{\|\mathbf{g}_l\|_2,1\}}$
		\EndParFor
		\EndFor
		\Ensure $\mathbf{X}^{(t_0)},\mathbf{D}^{(t_0)},\mathbf{Z}^{(t_0)}$.
	\end{algorithmic}
\end{algorithm}

\subsection{Convergence Analysis}
Inherited from the majorization-minimization framework, the non-increasing property \eqref{eq2.4} holds for any surrogate problem. And the objective value is bounded below by $0$ for problem \eqref{eq2.6}. So the sequence $\{f(\mathbf{x}^{(k)})\}$ generated by C-PRIME is guaranteed to converge to a finite point at least. Similarly, the sequence $\{f(\mathbf{X}^{(k)},\mathbf{D}^{(k)},\mathbf{Z}^{(k)})\}$ generated by SC-PRIME will also converge to a finite point since the block successive upper-bound minimization method (BSUM) also maintains the non-increasing property, and the objective value in problem \eqref{eq3.17} is also bounded below by $0$. Actually the BSUM method includes the majorization-minimization method as a special case by setting the majorization function as the upper-bound and the entire variable as one block. Therefore, it is sufficient to only study the convergence of the BSUM method, on which a thorough analysis can be found in \cite{Razaviyayn2013}. The authors have proven that every limit point of the iterates generated by the BSUM algorithm is a stationary point of the original problem if the constraint set is closed and convex (Algorithm 1 \cite{Razaviyayn2013}\footnote{The constraint set is limited to be in real space in \cite{Razaviyayn2013} to guarantee	differentiability. By introducing new variable $\tilde{\mathbf{x}}=[\mathrm{Re}(\mathbf{x})^T,\mathrm{Im}(\mathbf{x})^T]^T\in\mathbb{R}^{2N}$, it is easy to transform the problems in this parer into equivalent problems with real variables.}). Fortunately, \eqref{eq2.6} is an unconstrained optimization problem and \eqref{eq3.17} involves only one closed convex set. Hence the sequences of points generated by C-PRIME and SC-PRIME will converge to a stationary point of their corresponding original problems.

\subsection{Computational Complexity}
The updating procedures of our two algorithms are quite straightforward, only requiring basic matrix and vector operations. To recover a signal that is sparse in the standard basis, C-PRIME has a time complexity $O(MN)$ under a general measurement matrix setting and $O(M\log M)$ under a DFT measurement matrix setting by exploiting fast Fourier transform and inverse fast Fourier transform. When the unknown signal is not sparse, SC-PRIME utilizes the sparse coding technique to approximate the unknown signal by a linear combination of a few columns in a dictionary. The time complexity of SC-PRIME is $O(LNP)$ to solve the undersampled phase retrieval task.

\section{Numerical Results}
\label{section4}
In this section, we investigate the numerical performance of our MM-based algorithms, C-PRIME and SC-PRIME, and compare them with two up-to-date benchmark methods: UPRwO \cite{Weller2015} and DOLPHIn \cite{Tillmann2016}, respectively. First, we compare C-PRIME with UPRwO on the same randomly generated data. Later, we compare SC-PRIME with DOLPHIn on practical test images. Experimental results validate that C-PRIME and SC-PRIME outperform their corresponding benchmark method in terms of successful recovery probability and accuracy. All experiments are conducted on a personal computer with a $3.20$ GHz Intel Core i$5$-$4570$ CPU and $8.00$ GB RAM running Matlab R$2014$b.

\subsection{C-PRIME vs. UPRwO}
We first investigate the performance of C-PRIME and compare it with the benchmark method UPRwO \cite{Weller2015}. To implement the UPRwO algorithm, we use the code provided on the authors' homepage\footnote{\url{http://people.virginia.edu/~dsw8c/sw.html}}. In this subsection, we consider the clean measurements case and therefore we set the number of outliers to $0$ and signal-to-noise ratio (SNR) to infinity in the code. All other parameters are set as the default value.

The initialization steps of the UPRwO method are summarized below:
\begin{enumerate}
	\item Generating a random original signal $\mathbf{x}_o\in\mathbb{C}^{N}$ with cardinality $K$, where the support is also randomly selected;
	\item Forming the measurement matrix $\mathbf{A}\in\mathbb{C}^{M\times N}$ by randomly selecting $M$ rows in the $N\times N$ unitary DFT matrix (in this way $\mathbf{AA}^H=\mathbf{I}_M$ and $\mathbf{I}_N\succeq\mathbf{A}^H\mathbf{A}$);
	\item Generating the intensity measurements $\mathbf{y}=\left|\mathbf{Ax}_o\right|^2$.
\end{enumerate}
For a fair comparison, in the simulation, we run the UPRwO code first with a fixed $(N,M,K)$ value. Besides the final results, we also store the original signal $\mathbf{x}_o$, the measurement matrix $\mathbf{A}$, and the intensity measurements $\mathbf{y}$. Later, we run our C-PRIME code using the same measurement matrix $\mathbf{A}$ and intensity measurements $\mathbf{y}$ from the UPRwO simulation.

In detail, the length of the original signal $N$ is set as the default value $128$. Since we consider the undersampled phase retrieval problem, the number of measurements is limited to be $M\in\{128,64,32,16,8\}$ and the sparsity level is set to be $K\in\{16,8,4,2\}$ (a value larger than $16$ ends up with unsuccessful recovery). For each of these possible $(M,K)$ pairs, experiments are conducted to test the performance of both algorithms provided with the same measurement matrix and intensity measurements. Note that under the DFT measurement matrix setting, any individual or combination of the following three trivial ambiguities conserve the Fourier magnitude:
\begin{enumerate}
	\item Global constant phase shift: $\mathbf{x}\rightarrow\mathbf{x}\cdot e^{j\phi}$,
	\item Circular shift: $[\mathbf{x}]_i\rightarrow[\mathbf{x}]_{(i+i_0)\mod N}$,
	\item Conjugate inversion: $[\mathbf{x}]_i\rightarrow\overline{[\mathbf{x}]_{N-i}}$.
\end{enumerate}
Hence a disambiguation step is necessary to find the unique solution. For each solution $\mathbf{x}^{\star}$ returned by UPRwO and C-PRIME, we check all the possible candidates within the trivial ambiguities and choose the one with least normalized squared error (NSE) with regard to the original signal $\mathbf{x}_o$ as the final solution. The NSE between $\mathbf{x}^{\star}$ and $\mathbf{x}_o$ is calculated as
\begin{equation}
\text{NSE}(\mathbf{x}^{\star},\mathbf{x}_o)=\underset{\mathbf{x}\in\mathcal{S}(\mathbf{x}^{\star})}{\min}\frac{\|\mathbf{x}-\mathbf{x}_o\|_2^2}{\|\mathbf{x}_o\|_2^2},
\end{equation}
where the set $\mathcal{S}(\mathbf{x}^{\star})$ contains all the possible signals within the trivial ambiguities of $\mathbf{x}^{\star}$. Furthermore, since the original signal is generated as a random vector, experiments are repeated $100$ times for every $(M,K)$ pair using different and independent original signals with everything else fixed. The normalized mean squared error is calculated as the average of these $100$ independent NSE values. And among these $100$ independent trials, an algorithm is considered to successfully recover the original signal if the corresponding NSE is less than $10^{-4}$.

\begin{figure*}[t]
	\centering
	\subfigure[UPRwO]{\includegraphics[width=9cm]{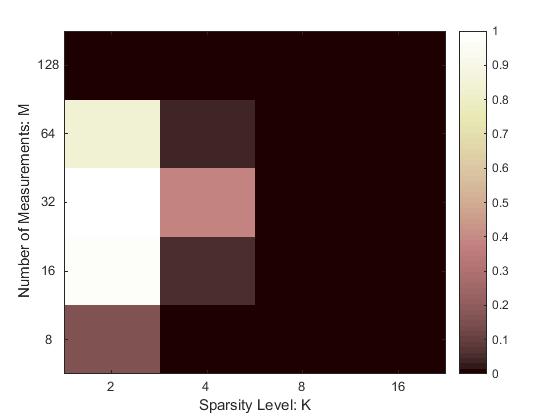}}
	\subfigure[C-PRIME]{\includegraphics[width=9cm]{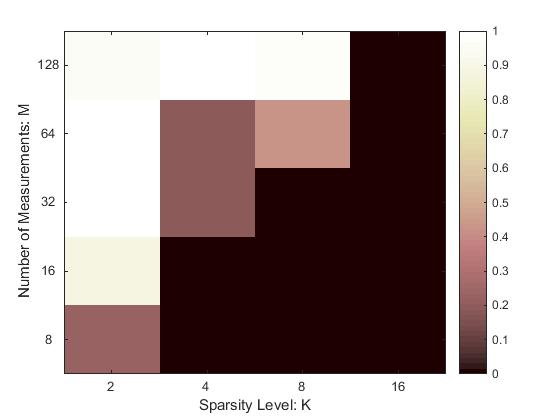}}
	\caption{Successful recovery probability of UPRwO and C-PRIME for an $N=128$-length complex signal, versus signal sparsity level $K$ and number of clean measurements $M$.}
	\label{fig_pro}
\end{figure*}

\begin{figure*}[t]
	\centering
	\subfigure[UPRwO]{\includegraphics[width=9cm]{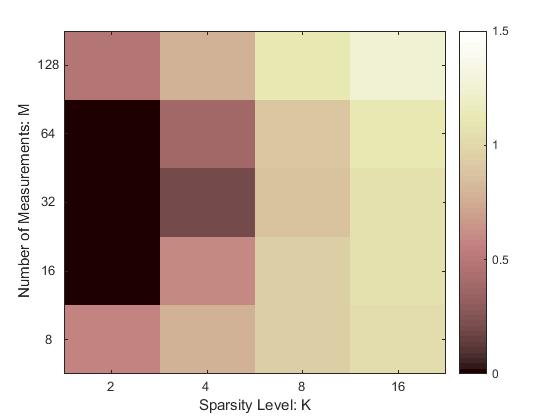}}
	\subfigure[C-PRIME]{\includegraphics[width=9cm]{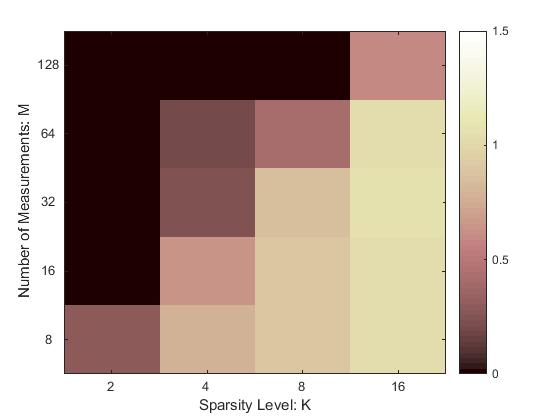}}
	\caption{Normalized mean squared error of UPRwO and C-PRIME for an $N=128$-length complex signal, versus signal sparsity level $K$ and number of clean measurements $M$.}
	\label{fig_nmse}
\end{figure*}

Final experimental results of UPRwO and C-PRIME are plotted in Fig. \ref{fig_pro} on the successful recovery probabilities, and Fig. \ref{fig_nmse} on the normalized mean squared error. For sparse signals under different $(M,K)$ settings, our MM-based algorithm C-PRIME has a larger successful recovery probability and less normalized mean squared error than the benchmark algorithm UPRwO. The average CPU times of UPRwO and C-PRIME over these $100$ independent trials under all $(M,K)$ settings are presented in Table \ref{table_cputime_UPRwO}. Both algorithms have a similar computational time.

\begin{table}[H]
	\renewcommand{\arraystretch}{1.3}
	\caption{Average CPU time of UPRwO and C-PRIME under different $(M,K)$ settings. The value is presented as ($t_\text{UPRwO}$, $t_\text{C-PRIME}$).}
	\label{table_cputime_UPRwO}
	\centering
	\begin{tabular}{|c|c|cccc|}
		\hline
		\multirow{5}{*}{$M$}&128&58.1, 135&57.0, 81.9&55.6, 127&57.7, 7.05\\
		&64&33.2, 39.6&33.1, 41.2&33.7, 40.5&33.3, 2.37\\
		&32&31.5, 27.4&32.5, 30.0&31.8, 30.3&30.5, 1.82\\
		&16&27.4, 6.84&30.3, 24.3&30.3, 24.3&29.2, 1.57\\
		&8&21.3, 18.7&25.7, 22.8&26.3, 23.0&25.2, 2.85\\ \hline
		\multicolumn{2}{|c|}{\multirow{2}{*}{Time(s)}}&2&4&8&16\\ \cline{3-6}
		\multicolumn{2}{|c|}{}&\multicolumn{4}{|c|}{$K$}\\ \hline
	\end{tabular} 
\end{table}

\begin{figure*}[t]
	\subfigure[Original]{\includegraphics[width=6cm]{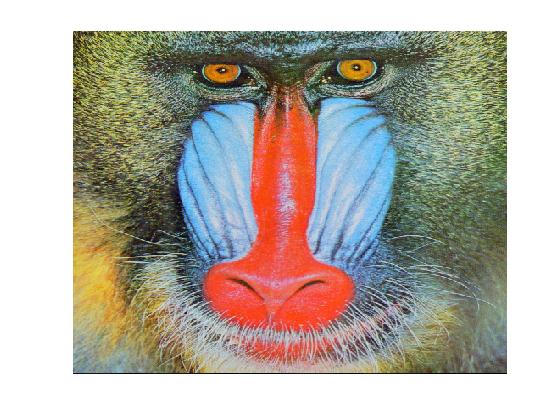}}
	\subfigure[DOLPHIn]{\includegraphics[width=6cm]{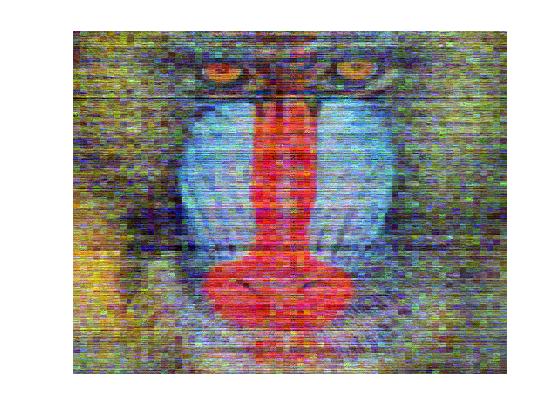}}
	\subfigure[SC-PRIME]{\includegraphics[width=6cm]{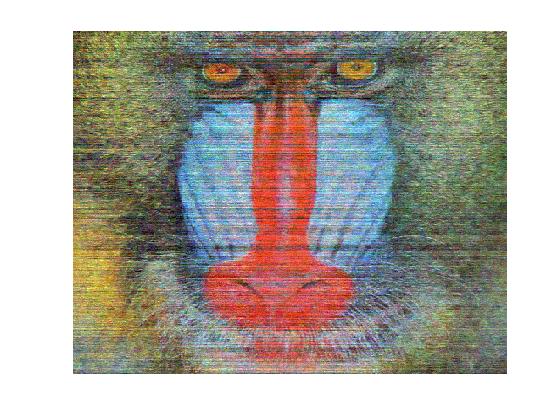}}
	\caption{Reconstruction results of DOLPHIn and SC-PRIME on the $512\times512$ color mandrill image. (a) the original image; (b) image reconstructed by DOLPHIn, PSNR $=14.84$ dB, SSIM $=0.4148$, t $=127.0$ s; (c) image reconstructed by SC-PRIME, PSNR $=17.17$ dB, SSIM $=0.5773$, t $=76.79$ s.}
	\label{fig_mandrill}
\end{figure*}

\begin{figure*}[t]
	\subfigure[Original]{\includegraphics[width=6cm]{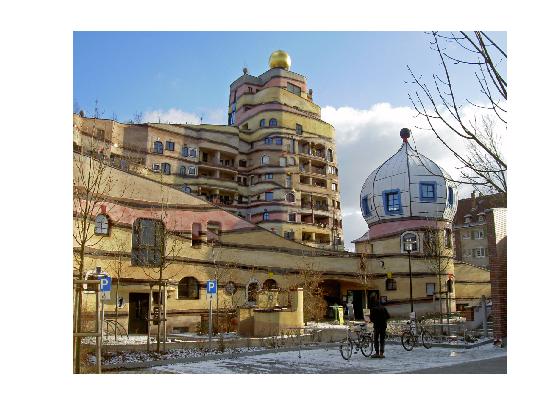}}
	\subfigure[DOLPHIn]{\includegraphics[width=6cm]{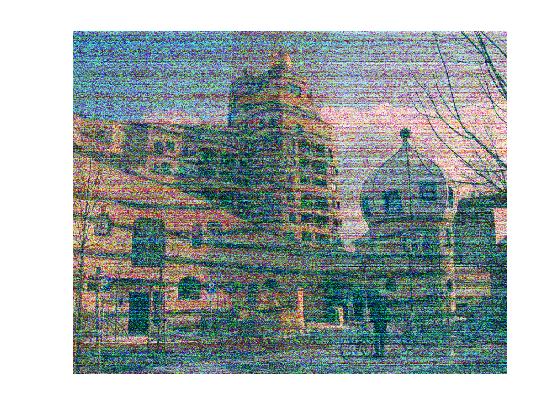}}
	\subfigure[SC-PRIME]{\includegraphics[width=6cm]{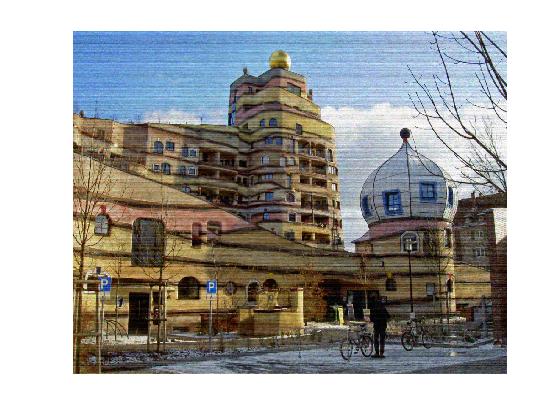}}
	\caption{Reconstruction results of DOLPHIn and SC-PRIME on the $2816\times2112$ color waldspirale image. (a) the original image; (b) image reconstructed by DOLPHIn, PSNR $=10.81$ dB, SSIM $=0.1143$, t $=3142$ s; (c) image reconstructed by SC-PRIME, PSNR $=20.60$ dB, SSIM $=0.6583$, t $=1442$ s.}
	\label{fig_waldspirale}
\end{figure*}

\subsection{SC-PRIME vs. DOLPHIn}
We now investigate the performance of SC-PRIME and compare it with the benchmark method DOLPHIn \cite{Tillmann2016} on practical test images. To implement DOLPHIn, we use the code provided on the authors' homepage\footnote{\url{http://www.mathematik.tu-darmstadt.de/~tillmann/\#software}}. The test images are also downloaded from the same website. We choose the Gaussian measurement matrix setting and change the sampling rate from $4$ to $0.5$ ($M=0.5N$) to set up a valid undersampled phase retrieval problem. All other parameters are kept as the default value.

At the initialization step, the DOLPHIn algorithm takes the $2$D image as the original signal, thereby generating a random complex Gaussian measurement matrix, and generating the noisy intensity measurements with additive white Gaussian noise. The default SNR is $15$ dB. First, we run the DOLPHIn code and store the measurement matrix as well as the noisy intensity measurements. Later, these same noisy intensity measurements and the measurement matrix are provided as the input for SC-PRIME. Different from our problem setting in Section III where the patch signals are directly treated as the target signal, there is a sorting step between the $2$D image signal and the $2$D patch-based signal in \cite{Tillmann2016}. The corresponding change in the implementation of SC-PRIME is easy and trivial since changing the order of the elements in a matrix conserves its Frobenius norm.

To evaluate the quality of the reconstructed images, two standard image quality metrics are considered in this paper, namely the peak signal-to-noise ratio (PSNR) and the structural similarity index (SSIM). PSNR is the ratio between the maximum possible power of the original image and the mean squared error between the reconstructed image and the original image. It is usually expressed in terms of the logarithmic decibel scale, and the larger the value the better the image quality. SSIM reflects the structural similarities between the reconstructed image and the original image. It is on a scale from $0$ to $1$, and a larger value represents more similarities in the structure to the original image.

Final reconstruction results of DOLPHIn and SC-PRIME are presented in Fig. \ref{fig_mandrill} on the $512\times 512$ color mandrill image, and Fig. \ref{fig_waldspirale} on the $2816\times2112$ color waldspirale image. In both cases, our MM-based algorithm SC-PRIME can reconstruct the image with a larger PSNR and SSIM value as well as an impressively better visual quality than the benchmark algorithm DOLPHIn. Moreover, we summarize in Table \ref{table_result} the PSNR and SSIM value of the reconstructed images for both algorithms on the rest of the test images. Besides the results of the reconstructed images ($\mathbf{X}^{\star}$), we also show the results of images approximated by the dictionary ($\mathbf{D}^{\star}\mathbf{Z}^{\star}$). All numbers in the table are averaged over $100$ Monte Carlo simulations using different and independent additive white Gaussian noise. The average CPU times over these $100$ independent trials are presented in Table \ref{table_cputime_DOLPHIn}.  It is interesting that the images approximated by the dictionary have a slightly larger PSNR and SSIM value than those reconstructed directly by the algorithms. Nevertheless, our MM-based algorithm SC-PRIME outperforms the benchmark method DOLPHIn in terms of PSNR and SSIM on all of the test images at the cost of slightly more CPU time. 
  
\begin{table*}[t]
	\renewcommand{\arraystretch}{1.3}
	\caption{Reconstruction results of DOLPHIn and SC-PRIME on different test images.}
	\label{table_result}
	\centering
	\begin{tabular}{|cr|c|c|c|c|c|c|c|c|}
		\hline
		\multicolumn{2}{|c|}{\multirow{2}{*}{PSNR(dB), SSIM}}&\multicolumn{3}{|c|}{$256\times256$ images}&\multicolumn{5}{|c|}{$512\times512$ images}\\ \cline{3-10}
		\multicolumn{2}{|c|}{}&Cameraman&House&Peppers&Barbara&Boat&Fingerprint&Lena&Mandrill\\ \hline 
		\multirow{2}{*}{DOLPHIn}&$\mathbf{X}$&16.2, 0.255&16.3, 0.235&16.1, 0.286&17.0, 0.291&17.1, 0.263&15.4, 0.394&17.5, 0.270&16.8, 0.282\\
		&$\mathbf{DZ}$&15.8, 0.266&15.9, 0.281&15.7, 0.308&16.4, 0.287&16.6, 0.281&15.1, 0.366&16.9, 0.319 &16.2, 0.256\\ \hline 
		\multirow{2}{*}{SC-PRIME} &$\mathbf{X}$&16.6, 0.296&18.0, 0.312&17.3, 0.344&18.3, 0.319&18.8, 0.285&16.4, 0.433&19.3, 0.284&19.2, 0.371\\ 
		&$\mathbf{DZ}$&16.0, 0.306&17.2, 0.387&16.5, 0.377&18.6, 0.359&19.0, 0.351&16.6, 0.446&19.7, 0.391&19.2, 0.390\\ 
		\hline 
	\end{tabular}
\end{table*}

\begin{table*}[t]
	\renewcommand{\arraystretch}{1.3}
	\caption{Average CPU time of DOLPHIn and SC-PRIME on different test images.}
	\label{table_cputime_DOLPHIn}
	\centering
	\begin{tabular}{|c|ccc|ccccc|}
		\hline
		\multirow{2}{*}{Time(s)}&\multicolumn{3}{|c|}{$256\times256$ images}&\multicolumn{5}{|c|}{$512\times512$ images}\\ \cline{2-9}
		&Cameraman &House&Peppers&Barbara&Boat&Fingerprint&Lena&Mandrill\\ \hline 
		DOLPHIn&7.97&8.07&8.05&22.3&21.0&22.6&20.1&23.4\\ 
		SC-PRIME&8.02&8.17&8.28&24.3&23.6&24.6&23.6&23.5\\ \hline 
	\end{tabular} 
\end{table*}

\section{Conclusion}
\label{section5}
The undersampled phase retrieval problem draws great attention in various imaging applications. The difficulty lies in both theoretical analysis and practical algorithm design. Provided that only undersampled intensity measurements are available, one needs to solve a non-linear, non-convex, and under-determined inverse problem. In this paper, we have proposed two efficient algorithms based on the majorization-minimization framework that outperform the up-to-date benchmark methods in terms of successful recovery probability and accuracy under different problem settings. The first algorithm C-PRIME can uniquely recover a signal that is sparse in the standard basis. When the signal is not sparse itself, the second algorithm SC-PRIME utilizes the sparse coding technique to approximate the target signal by a linear combination of a few columns in a dictionary. Experimental results on randomly generated data and practical test images are also provided in the paper to further validate the efficiency of our algorithms with impressive results.

\appendices
\section{Justification for Using Modulus Information}
Recall the $M$ noisy intensity measurements
\begin{equation}
y_i=\left|\mathbf{a}_i^H\mathbf{x}\right|^2+n_i,\;i=1,\ldots,M,
\end{equation}
and we assume $y_i\geq0$ (otherwise we just discard this measurement). And the noise $n_i$ is assumed to be independent of the measurements. Therefore,
\begin{equation}
\sqrt{y_i}=\sqrt{\left|\mathbf{a}_i^H\mathbf{x}\right|^2+n_i}=\left|\mathbf{a}_i^H\mathbf{x}\right|\sqrt{1+\frac{n_i}{\left|\mathbf{a}_i^H\mathbf{x}\right|^2}},\;\forall i.
\end{equation}
Usually the noise level is much smaller than the value of the clean intensity measurements, $|n_i|\ll\left|\mathbf{a}_i^H\mathbf{x}\right|^2$. It is sufficient to make the following approximation taking the first two terms in the Taylor series:
\begin{equation}
\sqrt{1+\frac{n_i}{\left|\mathbf{a}_i^H\mathbf{x}\right|^2}}\approx1+\frac{n_i}{2\left|\mathbf{a}_i^H\mathbf{x}\right|^2},\;\forall i.
\end{equation}
Hence,
\begin{equation}
\sqrt{y_i}\approx\left|\mathbf{a}_i^H\mathbf{x}\right|+\frac{n_i}{2\left|\mathbf{a}_i^H\mathbf{x}\right|},\;\forall i.
\end{equation}
The first term $\left|\mathbf{a}_i^H\mathbf{x}\right|$ is the actual clean modulus measurement. And the second term can be regarded as the additive noise, with expectation and variance
\begin{equation}
\begin{aligned}
\mathbb{E}\left[\frac{n_i}{2\left|\mathbf{a}_i^H\mathbf{x}\right|}\right]&=\frac{\mathbb{E}\left[n_i\right]}{2\left|\mathbf{a}_i^H\mathbf{x}\right|},\;\forall i,\\
\mathrm{Var}\left[\frac{n_i}{2\left|\mathbf{a}_i^H\mathbf{x}\right|}\right]&=\frac{\mathrm{Var}\left[n_i\right]}{4\left|\mathbf{a}_i^H\mathbf{x}\right|^2},\;\forall i.
\end{aligned}
\end{equation}
Therefore, the additive noise to the modulus information $\left|\mathbf{a}_i^H\mathbf{x}\right|$ has a lesser expectation and variance value than the additive noise to the intensity information $\left|\mathbf{a}_i^H\mathbf{x}\right|^2$ when $\left|\mathbf{a}_i^H\mathbf{x}\right|>\frac{1}{2}$.

\section{Proof of $\lambda_{\max}\left((\mathbf{D}^{(k)})^H\mathbf{D}^{(k)}\right)\leq L$}
Note that the dictionary $\mathbf{D}^{(k)}=[\mathbf{d}_1^{(k)},\ldots,\mathbf{d}_L^{(k)}]$ satisfies $\|\mathbf{d}_l^{(k)}\|_2\leq1,\forall l=1,\ldots,L$, so
\begin{equation}
\begin{aligned}
&\lambda_{\max}\left((\mathbf{D}^{(k)})^H\mathbf{D}^{(k)}\right)=\lambda_{\max}\left(\mathbf{D}^{(k)}(\mathbf{D}^{(k)})^H\right)\\
=&\underset{\mathbf{t}\neq\mathbf{0}}{\max}\frac{\mathbf{t}^H\mathbf{D}^{(k)}(\mathbf{D}^{(k)})^H\mathbf{t}}{\|\mathbf{t}\|_2^2}=\underset{\mathbf{t}\neq\mathbf{0}}{\max}\sum\limits_{l=1}^{L}\frac{\left|\mathbf{t}^H\mathbf{d}_l^{(k)}\right|^2}{\|\mathbf{t}\|_2^2}\\
\leq&\underset{\mathbf{t}\neq\mathbf{0}}{\max}\sum\limits_{l=1}^{L}\frac{\|\mathbf{t}\|_2^2\|\mathbf{d}_l^{(k)}\|_2^2}{\|\mathbf{t}\|_2^2}=\sum\limits_{l=1}^{L}\|\mathbf{d}_l^{(k)}\|_2^2\leq L.
\end{aligned}
\end{equation}
The equality is achieved when all $\{\mathbf{d}_l^{(k)}\}$ lie on the same line and $\|\mathbf{d}_l^{(k)}\|_2=1,\forall l$.

\bibliographystyle{IEEETran}
\bibliography{../References}

\begin{thebibliography}{10}
\providecommand{\url}[1]{#1}
\csname url@samestyle\endcsname
\providecommand{\newblock}{\relax}
\providecommand{\bibinfo}[2]{#2}
\providecommand{\BIBentrySTDinterwordspacing}{\spaceskip=0pt\relax}
\providecommand{\BIBentryALTinterwordstretchfactor}{4}
\providecommand{\BIBentryALTinterwordspacing}{\spaceskip=\fontdimen2\font plus
\BIBentryALTinterwordstretchfactor\fontdimen3\font minus
  \fontdimen4\font\relax}
\providecommand{\BIBforeignlanguage}[2]{{%
\expandafter\ifx\csname l@#1\endcsname\relax
\typeout{** WARNING: IEEEtran.bst: No hyphenation pattern has been}%
\typeout{** loaded for the language `#1'. Using the pattern for}%
\typeout{** the default language instead.}%
\else
\language=\csname l@#1\endcsname
\fi
#2}}
\providecommand{\BIBdecl}{\relax}
\BIBdecl

\bibitem{Walther1963}
\BIBentryALTinterwordspacing
A.~Walther, ``The question of phase retrieval in optics,'' \emph{Optica Acta:
  International Journal of Optics}, vol.~10, no.~1, pp. 41--49, 1963. [Online].
  Available: \url{http://dx.doi.org/10.1080/713817747}
\BIBentrySTDinterwordspacing

\bibitem{Fienup1987}
C.~Fienup and J.~Dainty, ``Phase retrieval and image reconstruction for
  astronomy,'' \emph{Image Recovery: Theory and Application}, pp. 231--275,
  1987.

\bibitem{Harrison1993}
\BIBentryALTinterwordspacing
R.~W. Harrison, ``Phase problem in crystallography,'' \emph{Journal of the
  Optical Society of America A}, vol.~10, no.~5, pp. 1046--1055, May 1993.
  [Online]. Available:
  \url{http://josaa.osa.org/abstract.cfm?URI=josaa-10-5-1046}
\BIBentrySTDinterwordspacing

\bibitem{Miao2002}
\BIBentryALTinterwordspacing
J.~Miao, T.~Ishikawa, B.~Johnson, E.~H. Anderson, B.~Lai, and K.~O. Hodgson,
  ``High resolution 3{D} x-ray diffraction microscopy,'' \emph{Physical Review
  Letters}, vol.~89, p. 088303, Aug 2002. [Online]. Available:
  \url{http://link.aps.org/doi/10.1103/PhysRevLett.89.088303}
\BIBentrySTDinterwordspacing

\bibitem{Gerkmann2015}
T.~Gerkmann, M.~Krawczyk-Becker, and J.~Le~Roux, ``Phase processing for
  single-channel speech enhancement: History and recent advances,'' \emph{IEEE
  Signal Processing Magazine}, vol.~32, no.~2, pp. 55--66, March 2015.

\bibitem{Shechtman2015}
Y.~Shechtman, Y.~Eldar, O.~Cohen, H.~Chapman, J.~Miao, and M.~Segev, ``Phase
  retrieval with application to optical imaging: A contemporary overview,''
  \emph{IEEE Signal Processing Magazine}, vol.~32, no.~3, pp. 87--109, May
  2015.

\bibitem{Candes2013b}
E.~J. Cand{\`e}s, T.~Strohmer, and V.~Voroninski, ``Phaselift: Exact and stable
  signal recovery from magnitude measurements via convex programming,''
  \emph{Communications on Pure and Applied Mathematics}, vol.~66, no.~8, pp.
  1241--1274, 2013.

\bibitem{Heinosaari2013}
\BIBentryALTinterwordspacing
T.~Heinosaari, L.~Mazzarella, and M.~M. Wolf, ``Quantum tomography under prior
  information,'' \emph{Communications in Mathematical Physics}, vol. 318,
  no.~2, pp. 355--374, 2013. [Online]. Available:
  \url{http://dx.doi.org/10.1007/s00220-013-1671-8}
\BIBentrySTDinterwordspacing

\bibitem{Bodmann2015}
\BIBentryALTinterwordspacing
B.~G. Bodmann and N.~Hammen, ``Stable phase retrieval with low-redundancy
  frames,'' \emph{Advances in Computational Mathematics}, vol.~41, no.~2, pp.
  317--331, 2015. [Online]. Available:
  \url{http://dx.doi.org/10.1007/s10444-014-9359-y}
\BIBentrySTDinterwordspacing

\bibitem{Bandeira2014}
\BIBentryALTinterwordspacing
A.~S. Bandeira, J.~Cahill, D.~G. Mixon, and A.~A. Nelson, ``Saving phase:
  Injectivity and stability for phase retrieval,'' \emph{Applied and
  Computational Harmonic Analysis}, vol.~37, no.~1, pp. 106--125, 2014.
  [Online]. Available:
  \url{http://www.sciencedirect.com/science/article/pii/S1063520313000936}
\BIBentrySTDinterwordspacing

\bibitem{Qiu2016}
T.~Qiu, P.~Babu, and D.~P. Palomar, ``{PRIME}: Phase retrieval via
  {M}ajorization-{M}inimization,'' \emph{IEEE Transactions on Signal
  Processing}, vol.~64, no.~19, pp. 5174--5186, Oct 2016.

\bibitem{Balan2006}
\BIBentryALTinterwordspacing
R.~Balan, P.~Casazza, and D.~Edidin, ``On signal reconstruction without
  phase,'' \emph{Applied and Computational Harmonic Analysis}, vol.~20, no.~3,
  pp. 345--356, 2006. [Online]. Available:
  \url{http://www.sciencedirect.com/science/article/pii/S1063520305000667}
\BIBentrySTDinterwordspacing

\bibitem{Eldar2015}
Y.~C. Eldar, P.~Sidorenko, D.~G. Mixon, S.~Barel, and O.~Cohen, ``Sparse phase
  retrieval from short-time fourier measurements,'' \emph{IEEE Signal
  Processing Letters}, vol.~22, no.~5, pp. 638--642, May 2015.

\bibitem{Voroninski2016}
\BIBentryALTinterwordspacing
V.~Voroninski and Z.~Xu, ``A strong restricted isometry property, with an
  application to phaseless compressed sensing,'' \emph{Applied and
  Computational Harmonic Analysis}, vol.~40, no.~2, pp. 386--395, 2016.
  [Online]. Available:
  \url{http://www.sciencedirect.com/science/article/pii/S1063520315000901}
\BIBentrySTDinterwordspacing

\bibitem{Li2013}
\BIBentryALTinterwordspacing
X.~Li and V.~Voroninski, ``Sparse signal recovery from quadratic measurements
  via convex programming,'' \emph{SIAM Journal on Mathematical Analysis},
  vol.~45, no.~5, pp. 3019--3033, 2013. [Online]. Available:
  \url{http://dx.doi.org/10.1137/120893707}
\BIBentrySTDinterwordspacing

\bibitem{Shechtman2014}
Y.~Shechtman, A.~Beck, and Y.~Eldar, ``{GESPAR}: Efficient phase retrieval of
  sparse signals,'' \emph{IEEE Transactions on Signal Processing}, vol.~62,
  no.~4, pp. 928--938, Feb 2014.

\bibitem{Mukherjee2014}
S.~Mukherjee and C.~S. Seelamantula, ``Fienup algorithm with sparsity
  constraints: Application to frequency-domain optical-coherence tomography,''
  \emph{IEEE Transactions on Signal Processing}, vol.~62, no.~18, pp.
  4659--4672, Sept 2014.

\bibitem{Weller2015}
D.~Weller, A.~Pnueli, G.~Divon, O.~Radzyner, Y.~Eldar, and J.~Fessler,
  ``Undersampled phase retrieval with outliers,'' \emph{IEEE Transactions on
  Computational Imaging}, vol.~1, no.~4, pp. 247--258, Dec 2015.

\bibitem{Moravec2007}
\BIBentryALTinterwordspacing
M.~L. Moravec, J.~K. Romberg, and R.~G. Baraniuk, ``Compressive phase
  retrieval,'' \emph{Proc. SPIE}, vol. 6701, pp. 670\,120--670\,120--11, 2007.
  [Online]. Available: \url{http://dx.doi.org/10.1117/12.736360}
\BIBentrySTDinterwordspacing

\bibitem{Fienup1982}
J.~R. Fienup, ``Phase retrieval algorithms: a comparison,'' \emph{Applied
  optics}, vol.~21, no.~15, pp. 2758--2769, 1982.

\bibitem{Ohlsson2012}
H.~Ohlsson, A.~Yang, R.~Dong, and S.~Sastry, ``{CPRL} -- {A}n extension of
  compressive sensing to the phase retrieval problem,'' in \emph{Advances in
  Neural Information Processing Systems 25}, F.~Pereira, C.~Burges, L.~Bottou,
  and K.~Weinberger, Eds.\hskip 1em plus 0.5em minus 0.4em\relax Curran
  Associates, Inc., 2012, pp. 1367--1375.

\bibitem{Schniter2015}
P.~Schniter and S.~Rangan, ``Compressive phase retrieval via generalized
  approximate message passing,'' \emph{IEEE Transactions on Signal Processing},
  vol.~63, no.~4, pp. 1043--1055, Feb 2015.

\bibitem{Tillmann2016}
A.~M. Tillmann, Y.~C. Eldar, and J.~Mairal, ``{DOLPHI}n-{D}ictionary learning
  for phase retrieval,'' \emph{arXiv preprint arXiv:1602.02263}, 2016.

\bibitem{Aharon2006}
M.~Aharon, M.~Elad, and A.~Bruckstein, ``\rm{K}-{SVD}: An algorithm for
  designing overcomplete dictionaries for sparse representation,'' \emph{IEEE
  Transactions on Signal Processing}, vol.~54, no.~11, pp. 4311--4322, Nov
  2006.

\bibitem{Elad2006}
M.~Elad and M.~Aharon, ``Image denoising via sparse and redundant
  representations over learned dictionaries,'' \emph{IEEE Transactions on Image
  Processing}, vol.~15, no.~12, pp. 3736--3745, Dec 2006.

\bibitem{Mairal2010}
\BIBentryALTinterwordspacing
J.~Mairal, F.~Bach, J.~Ponce, and G.~Sapiro, ``Online learning for matrix
  factorization and sparse coding,'' \emph{The Journal of Machine Learning
  Research}, vol.~11, pp. 19--60, Mar. 2010. [Online]. Available:
  \url{http://dl.acm.org/citation.cfm?id=1756006.1756008}
\BIBentrySTDinterwordspacing

\bibitem{Mairal2014}
\BIBentryALTinterwordspacing
J.~Mairal, F.~Bach, and J.~Ponce, ``Sparse modeling for image and vision
  processing,'' \emph{Foundations and Trends in Computer Graphics and Vision},
  vol.~8, no. 2-3, pp. 85--283, 2014. [Online]. Available:
  \url{http://dx.doi.org/10.1561/0600000058}
\BIBentrySTDinterwordspacing

\bibitem{Hunter2004}
\BIBentryALTinterwordspacing
D.~R. Hunter and K.~Lange, ``A tutorial on {MM} algorithms,'' \emph{The
  American Statistician}, vol.~58, no.~1, pp. 30--37, 2004. [Online].
  Available: \url{http://dx.doi.org/10.1198/0003130042836}
\BIBentrySTDinterwordspacing

\bibitem{Sun2016}
Y.~Sun, P.~Babu, and D.~P. Palomar, ``Majorization-minimization algorithms in
  signal processing, communications, and machine learning,'' \emph{IEEE
  Transactions on Signal Processing}, 2016.

\bibitem{Song2014}
J.~Song, P.~Babu, and D.~P. Palomar, ``Optimization methods for designing
  sequences with low autocorrelation sidelobes,'' \emph{IEEE Transactions on
  Signal Processing}, vol.~63, no.~15, pp. 3998--4009, Aug 2015.

\bibitem{VARADHAN2008}
\BIBentryALTinterwordspacing
R.~Varadhan and C.~Roland, ``Simple and globally convergent methods for
  accelerating the convergence of any {EM} algorithm,'' \emph{Scandinavian
  Journal of Statistics}, vol.~35, no.~2, pp. 335--353, 2008. [Online].
  Available: \url{http://dx.doi.org/10.1111/j.1467-9469.2007.00585.x}
\BIBentrySTDinterwordspacing

\bibitem{Hong2016}
M.~Hong, M.~Razaviyayn, Z.~Q. Luo, and J.~S. Pang, ``A unified algorithmic
  framework for block-structured optimization involving big data: With
  applications in machine learning and signal processing,'' \emph{IEEE Signal
  Processing Magazine}, vol.~33, no.~1, pp. 57--77, Jan 2016.

\bibitem{Razaviyayn2013}
\BIBentryALTinterwordspacing
M.~Razaviyayn, M.~Hong, and Z.-Q. Luo, ``A unified convergence analysis of
  block successive minimization methods for nonsmooth optimization,''
  \emph{SIAM Journal on Optimization}, vol.~23, no.~2, pp. 1126--1153, 2013.
  [Online]. Available: \url{http://dx.doi.org/10.1137/120891009}
\BIBentrySTDinterwordspacing

\end{thebibliography}





\end{document}